\documentclass[twocolumn,showpacs,amsmath,prl]{revtex4}
\usepackage{graphicx}
\usepackage{dcolumn}
\usepackage{bm}

\begin{document}

\title{Electronic Transport in the Oxygen Deficient \\ 
Ferromagnetic Semiconducting TiO$_{2-\delta}$}

\author{Soack Dae Yoon}
\altaffiliation[Corresponding author e-mail: syoon@ece.neu.edu ]{}
\author{Vincent G. Harris}
\email{harris@ece.neu.edu}
\author{Carmine Vittoria}
\email{vittoria@neu.edu}
\affiliation{Center for Microwave Magnetic Materials 
and Integrated Circuits, Department of Electrical and 
Computer Engineering, Northeastern University, 
Boston, MA. 02115 USA}

\author{Allan Widom}
\email{allan.widom@gmail.com}
\affiliation{Department of Physics, Northeastern University, 
Boston, MA. 02115 USA}%

\begin{abstract}
TiO$_{2-\delta}$ films were deposited on (100) Lanthanum 
aluminates LaAlO$_{3}$ substrates at a very low oxygen chamber 
pressure $P\approx 0.3$ mtorr employing a pulsed laser ablation 
deposition technique. In previous work, it was established that 
the oxygen deficiency in these films induced ferromagnetism. In this 
work it is demonstrated that this same oxygen deficiency also gives 
rise to semiconductor titanium ion impurity donor energy 
levels. Transport resistivity measurements in thin films of 
TiO$_{2-\delta}$ are presented as a function of temperature and 
magnetic field. Magneto- and Hall- resistivity is explained in 
terms of electronic excitations from the titanium ion donor levels  
into the conduction band. 
\end{abstract}


\maketitle

\section{introduction}
\label{intro}

Today, material research is focused toward a technology base whereby 
miniature devices operating at high speed, over a wide range of 
frequencies and exhibiting multifunctional properties can be developed 
routinely and efficiently\cite{Ohno:1998,Prinz:1998,Wolf:2001}. Titanium 
dioxide TiO$_2$ is a well known wide band gap oxide semiconductor  
belonging to the group IV-VI semiconductors. It is described in terms of a 
model structure of Ti$^{4+}$ and O$^{2-}$ ions. It has large dielectric 
constant and dielectric anisotropy with various crystal  
structures. TiO$_{2}$ is known to be an n-type semiconductor with a large 
energy gap in the range $3 {\rm \ volt}<\Delta /e< 9 {\rm \ volt}$ depending 
on sample preparation\cite{Masumoto:2001,Toyoaki:2003,Higgins:2004,Earle:1942,
Breckenridge:1953,Daude:1977,Pascual:1978,Tang:1993}. 
Previously\cite{Yoon:2006}, thin TiO$_{2-\delta}$ 
films of thickness in the range $200 {\rm \ nm}< t < 400 {\rm \ nm}$  
were deposited on substrates of (100) of lanthanum aluminate LaAlO$_{3}$. 
Bulk TiO$_{2}$ does not order magnetically at room temperature. However, 
spontaneous magnetization in TiO$_{2-\delta}$ films\cite{Yoon:2006}  
occurred in the temperature range $4 {\rm \ ^oK}< T < 880 {\rm \ ^oK}$.
The spontaneous magnetization was attributed to the oxygen deficiency  
induced during the growth of the films. It was argued that 
oxygen deficiency caused an imbalance in the ionic charge neutrality 
of the film thereby creating electron donor impurity ions of Ti$^{2+}$ and  
Ti$^{3+}$ in addition to the usual Ti$^{4+}$ ions. 
Ferromagnetism, i.e. net magnetic moments in the Ti$^{2+}$ and Ti$^{3+}$ ions, 
was a result of superexchange interactions between these ions via the oxygen 
ions. Calculations indicate that of all possible superexchange permutations,  
double exchange may be the most predominant interaction in view of the 
filled electron bands of the oxygen ions\cite{Zuo}.
It is well known\cite{Zener:1951,De_Gennes:1960} that such double exchange 
favors ferromagnetism. 

Our purpose is to argue that the {\em same} magnetic ions and oxygen 
deficiency that gave rise to magnetism in TiO$_{2-\delta}$ films 
also play a vital role in the electronic {\em transport properties}. 
Specifically, the existence of donor Ti$^{2+}$ and Ti$^{3+}$ donor ions 
in our films allow for electronic excitations into conduction bands. 
The number of carriers in the conduction band varies only mildly with 
magnetic field and more strongly with temperature. Indeed, the 
measurements reported here of the Hall resistivity is approximately 
linear in the magnetic field and the the magneto-resistivity is only very 
mildly field dependent. The magneto- and Hall- resistivity have very 
similar temperature dependence reflecting the variation of the number 
of mobile electron transport carriers with temperature. Measurements of 
transport properties of TiO$_{2}$ films as well as the theoretical analysis 
are discussed in the next section. The films were deposited employing a 
pulsed laser ablation deposition (PLD) technique at an oxygen pressure 
$P\approx 0.3$ mtorr. Final arguments are presented in the concluding 
section.

\section{Magneto-Transport Measurements}
\label{measure}

Pulsed laser ablation deposition (PLD) techniques were employed with a 
weakly paramagnetic magnetic TiO$_{2}$ target. Thin films of oxygen 
deficient TiO$_{2-\delta}$  were produced on (100) lanthanum aluminate 
LaAlO$_3$ substrates. During the PLD process, the substrate temperature 
was fixed at $T=700\ {\rm ^oC}$ while varying the oxygen pressure in the 
range $0.3\ {\rm mtorr}<P<400\ {\rm mtorr}$. Films deposited at 
$P\approx 0.3 {\rm mtorr}$ exhibited spontaneous magnetization and were 
therefore selected for measurements of magneto-transport film properties. 
The thickness of the films were in the range $200 {\rm \ nm}< t < 400 {\rm \ nm}$. 
The details of the magnetic and crystallographic characterization as well 
as other impurities and contaminants in the films were previously 
reported\cite{Yoon:2006}.

For steady currents and with the magnetic intensity ${\bf H}$ directed 
normal to the film, the resistance matrix ${\sf R}$ in the plane of the film 
may be written as\cite{Landau:1982} 
\begin{equation}
{\sf R}=
\begin{pmatrix}
R_{xx} & R_{xy} \\ 
R_{yx} & R_{yy}
\end{pmatrix}
=\frac{1}{t}
\begin{pmatrix}
\rho & -\rho_H \\ 
\rho_H & \rho
\end{pmatrix}
\label{transport1}
\end{equation}
wherein $\rho $ and $\rho_H$ represent, respectively, the magneto-resistivity 
and the Hall resistivity. The resistance matrix ${\sf R}$ of the TiO$_{2-\delta}$ 
films was measured by a conventional four-terminal technique using a gold-plated 
resistance sample puck from the quantum design physical property measurement 
system (PPMS). In that the ferromagnetic moment per unit volume is small on 
the scale of applied magnetic intensities, $|{\bf M}|\ll |{\bf H}|$, the magnetic 
field intensity may be identified with the magnetic induction 
${\bf B}=\mu_0({\bf H}+{\bf M})$, i.e. ${\bf B}\approx \mu_0{\bf H}$.
As a consequence of such a weak magnetization ${\bf M}$, the anomalous Hall 
resistivity is negligible. The original Hall expression for $\rho_H$ is sufficiently 
accurate. The magneto-resistivity is here described by the Drude model. Altogether, 
we may analyze the data in terms of the simple conventional expressions    
\begin{equation}
\rho_{H}=\frac{B}{ne}\ \ \ {\rm and}
\ \ \ \rho=\frac{m}{ne^2\tau }\ ,
\label{transport2}
\end{equation}
wherein $n$ is the density per unit volume of carriers in the conduction band.

\begin{figure}[tp]
\centering
\includegraphics[width=0.45\textwidth]{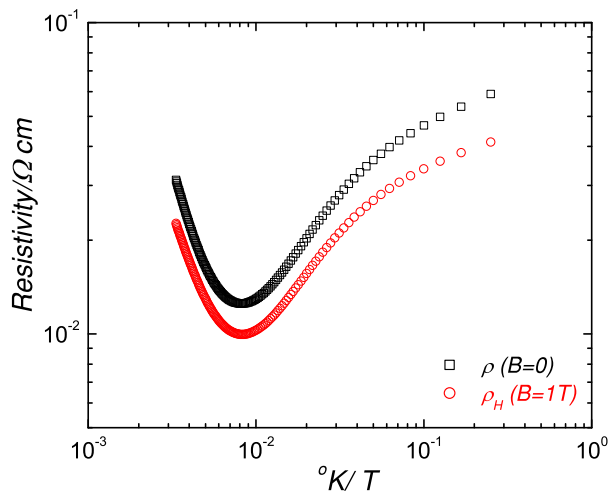}
\caption{Plotted on logarithmic scales are the magneto-resistivity $\rho $ 
at zero magnetic field intensity and the Hall resistivity $\rho_H$ 
at $B=1$ Tesla versus inverse temperature. The temperature variations are quite 
similar, i.e. the Hall angle $\theta_H=\arctan (\rho_H/\rho )$ is fairly uniform 
in temperature.}
\label{fig1} 
\end{figure}

In FIG. \ref{fig1} is shown measurements of $\rho $ at zero magnetic intensity 
and $\rho_H$ at $\mu_0 H=1.0$ Tesla is plotted as a function temperature $T$.
One notes that the temperature variations of $\rho $ and $\rho_H$ are quite similar 
as they are in other semiconductors\cite{Aschroft:1976,Debye:1954}  
The Hall angle $\theta_H$, as defined by   
\begin{equation}
\tan \theta_H=\frac{\rho_H}{\rho }\equiv \frac{eB\tau }{m}
\equiv \omega_c \tau\ ,
\label{transport3}
\end{equation}
is experimentally fairly uniform in temperature.

\begin{figure}[tp]
\centering
\includegraphics[width=0.45\textwidth]{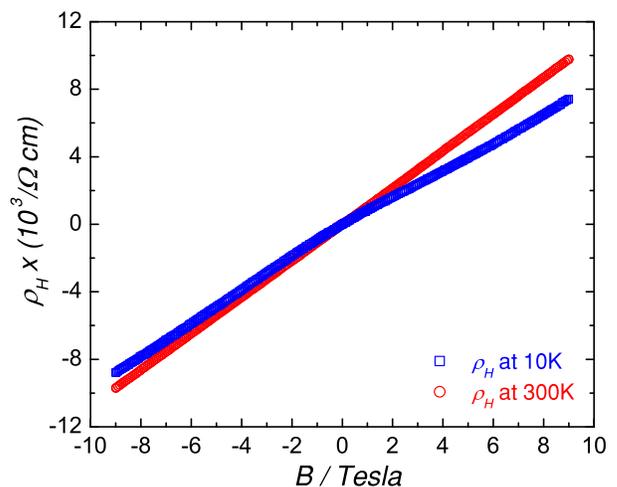}
\caption{ Shown are plots for a film deposited at 0.3 mtorr of the Hall resistivity 
$\rho_H$ for two temperatures, $T=10\ ^oK$ and $T=300\ ^oK$. In the high temperature regime, 
the linear behavior in magnetic field $\rho_H=(B/ne)$ is accurately obeyed. In the low 
temperature regime there are experimental deviations from linear behavior.}
\label{fig2} 
\end{figure}
\begin{figure}[tp]
\centering
\includegraphics[width=0.45\textwidth]{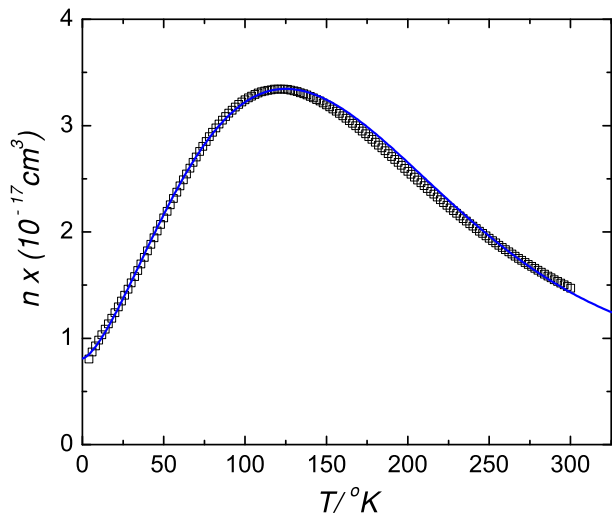}
\caption{Shown are the inferred density of carriers $n$ as a function of temperature 
taken from Hall resistivity data in a fixed magnetic field of $B=1$ Tesla. Also shown 
(solid curve) is the fit to Eqs.(\ref{transport5}) and (\ref{transport6}) with 
$\xi=0.0065$ and $\theta=125\ ^oK$.}
\label{fig3} 
\end{figure}

The hall resistivity $\rho_H$ as a function of magnetic field were measured   
at fixed temperature with applied field sweeps in the range 
$-9\ {\rm Tesla}<\mu_0 H <+9\ {\rm Tesla}$. The data for a high and low 
temperature are plotted in FIG. \ref{fig2}. For high temperatures, the experimental 
data is in excellent agreement with the linear magnetic field behavior for $\rho_H$ 
in Eq.(\ref{transport2}). For low temperatures, the agreement with the linear behavior 
in magnetic field is only fair. Nevertheless, it is experimentally clear that the
carriers are n-type and the carrier densities may be determined by 
\begin{equation}
n=\frac{B}{e\rho_H},
\label{transport4}
\end{equation}  
quite accurately for high temperatures and somewhat less accurately for low temperatures.
The resulting variation of the density of carriers $n$ with temperature $T$ is shown in 
FIG. \ref{fig3}.

The two striking features of the resulting inferred density of carriers are as follows: 
(i) The density of carriers does not appear to vanish in the low temperature limit 
$T\to 0$. In fact, there appears to be a low density of carriers in the ground state 
TiO$_{2-\delta }$ system of approximately 
\begin{equation}
n_0\approx \frac{8.1\times 10^{16}}{\rm cm^3}
\label{transport5}
\end{equation} 
at liquid helium temperatures.  This suggests that the electron chemical potential 
is placed very slightly into the conduction band by the Ti$^{2+}$ donors. The 
double exchange mechanism lowers that part of the conduction band containing 
electrons with spins parallel to those localized spins producing the 
magnetization\cite{Jungwirth:2006}. These effects allow for the low 
density of ground state carriers. (ii) The density of carriers as a function of 
temperature exhibits a clear maximum. We attribute 
this maximum in the number of carriers to the decrease of the magnetization $M$ 
to values below the saturation value $M_s$ as the temperature is raised. Theoretically, 
one finds from spin wave theory that $\lim_{T\to 0}[1-(M(T)/M_s)]/T^{3/2}=A_0$. For the 
temperature range here of experimental interest, we find that 
$M(T)\approx M_s[1-A_1T^{5/2}]$. If the rise in the carrier band lowest energy state is 
proportional to the the deviation of the magnetization from saturation\cite{Kubo:1974}, 
i.e. if $\Delta E\propto (M_s-M)$, then a reasonable phenomenological expression for the 
density of carriers, both in the ground state and in the thermally excited 
states, is  
\begin{equation}
n(T) = n_0+\frac{\xi }{\lambda_T^3}
\exp \left[-\left(\frac{T}{\theta }\right)^{3/2}\right]\ ,
\label{transport6}
\end{equation}
wherein the electron thermal wavelebgth
\begin{equation}
\lambda_T = \sqrt{\frac{2\pi \hbar^2}{mk_BT}}\ ,
\label{transport6a}
\end{equation}
$\xi $ is a dimensionless constant and $\theta $ is derived from 
the Boltzman factor $\exp(-\Delta E/k_BT)$ on the right hand side of 
Eq.(\ref{transport6}). That the density of carriers is the sum of ground 
state and thermally activated terms is at the heart of our phenomenology. 
As can be seen from the data in FIG. \ref{fig3}, the model gives rise to 
a good fit to the experimental data.

\begin{figure}[tp] 
\centering
\includegraphics[width=0.45\textwidth]{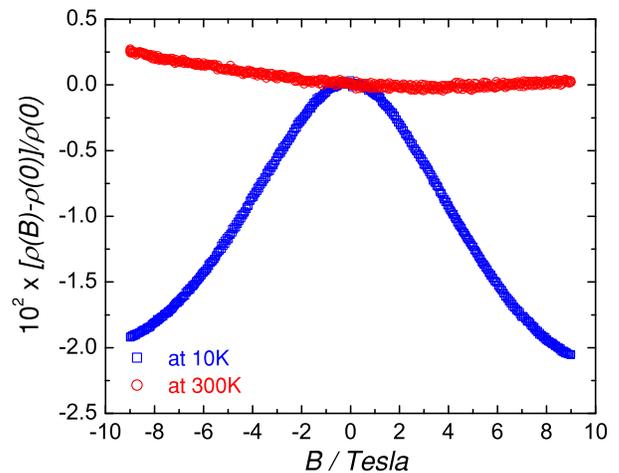}
\caption{The fractional change $f$ of the magneto-resistance in 
Eq.(\ref{transport7}) is plotted as the function of magnetic 
field. The blue squares and red circles represent, respectively, 
the measurements taken at temperatures $T=10\ ^oK $ and $300\ ^oK$.}
\label{fig4} 
\end{figure}

Finally in FIG. \ref{fig4}, the magneto-resistivity is plotted as a function of 
magnetic field in both the high and low temperature regimes. One may view the 
fractional change in magneto-resistivity, 
\begin{equation}
f(B,T)=\frac{\rho (B,T)-\rho (0,T)}{\rho (0,T)}\ ,
\label{transport7}
\end{equation} 
as a measure of how strongly the carrier density $n$ and the carrier lifetime 
$\tau $ vary with magnetic field. In the high temperature regime, the fractional 
change in the magneto-resistivity is quite small $|f_{high}|<0.003$.
In the low temperature regime the fractional change in magneto-resistivity is again 
fairly small $|f_{low}|<0.022$. Altogether, the magnetic field variations of the 
magneto-resistivity obeys $|f|<2.2\%$ in all regimes. That the deviations are 
small provides experimental support for the Hall-Drude model of electronic transport 
in Eq.(\ref{transport2}) for the oxygen deficient TiO$_{2-\delta }$  ferromagnetic 
semiconductor.

\section{conclusion}
\label{conc}

Measurements of the magneto- and Hall- resistivity have been measured in  
oxygen deficient TiO$_{2-\delta}$ magnetic semiconductor films. The data 
can be understood on the basis of the Hall-Drude model of resistivity 
from mobile carriers. The anomalous magnetic moment contribution to the Hall 
resistivity is small in most regimes since the ferromagnetic moment ${\bf }$ 
resides mainly on the dilutely distributed ions Ti$^{2+}$ and Ti$^{3+}$ and the 
magnitude is small $|{\bf M}|\ll |{\bf H}|$. The Drude relation time $\tau $ 
is only weakly dependent upon the magnetic field ${\bf B}$ and the temperature 
$T$ so that both the Hall resistivity $\rho_H$ and the normal resistivity 
$\rho $ vary similarly with temperature. The experimental data can thereby 
be described in terms of a temperature dependent mobile carrier density $n(T)$.
Apart from an expected Zeeman internal magnetic field splitting of the conduction 
band wherein the carrier spins align themselves parallel to the spins of the 
local ionic ferromagnetic moments, there also appears to be a small but finite 
mobile carrier density that persists even in the $T\to 0$ limit. The theory of such 
quantum ground state carriers in double ferromagnetic exchange semiconductors 
is worthy of further investigation.

\end{document}